# Conformable Fractional Bohr Hamiltonian with Bonatsos and Double-Well Sextic Potentials

M. M. Hammad

Faculty of Science, Department of Mathematics, Damanhour University, Egypt

**Abstract**

Using the conformable fractional calculus, a new formulation of the Bohr Hamiltonian is introduced. The conformable fractional energy spectra of free- and two- parameters anharmonic oscillator potentials are investigated. The energy eigenvalues and wave functions are calculated utilizing the finite-difference discretization method. It is proved that the conformable fractional spectra of the free-parameter Bonatsos potentials, $\beta^{2n}/2$, close completely the gaps between the classical spectra of the vibrational $U(5)$ dynamical symmetry, the $E(5) - \beta^{2n}$ models, and the $E(5)$ critical point symmetry. The ground effective sextic potential, which generates both the ground state and the $\beta$ excited states $0^+$, is considered to have two degenerate minima. In this case, the conformable fractional spectra of sextic potentials show a change, as a function of barrier height, from $\gamma$-unstable $O(6)$ energy level sequence to the spectrum of $E(5)-\beta^6$ model and simultaneously provide new features. The shape coexistence phenomena, in the ground band states, are identified. The energy spectrum and shape coexistence with mixing phenomena in $^{96}$Mo nucleus are discussed in the framework of the conformable fractional Bohr Hamiltonian.

## 1. Introduction

Several quantum systems are distinguished in their equilibrium state by a phase or shape. These shapes are, in numerous situations, rigid. Nevertheless, there are cases in which the system displays a phase transition between various phases. In atomic nuclei, the study of quantum shape phase transition (QSPT) [1-3] is complicated because of the finite number of particles that constitutes these systems. Usually, QSPT can be explained using the interacting boson model [4] and the collective model [5-8]. Three dynamical symmetries $U(5)$, $SU(3)$, and $O(6)$ in the interacting boson model are shown to correspond to three stable shapes of nuclei, namely, spherical, axially deformed, and γ-soft deformed shapes, respectively. The QSPTs coincide with transitions between dynamical symmetries, with a first- and second-order QSPTs taking place in the $U(5)$-$SU(3)$ and $U(5)$-$O(6)$ transitions, respectively. Moreover, the $X(5)$ [9] and $E(5)$ [10] critical point symmetries (CPSs) were introduced to specify the nuclei at the critical points corresponding to the $U(5)$-$SU(3)$ and $U(5)$-$O(6)$ transitions, respectively. Typically, these CPSs, that provide parameter-free predictions, come exactly or approximately using Bohr Hamiltonian (BH) with appropriate $\beta$- and $\gamma$-potentials depending on the physical cases studied. Particularly, in the $E(5)$ CPS, the potential was assumed to be a function only in the collective variable $\beta$. Subsequently, the exact separation of variables was accomplished and the analytic solution using an infinite square well potential (ISWP) in the equation containing the variable $\beta$ was obtained. The nuclei $^{102}$Pd, $^{104}$Ru, $^{106}$Cd, $^{106}$Mo, $^{108}$Cd, $^{124}$Te, $^{128}$Xe, and $^{134}$Ba have been identified as the empirical realizations of the $E(5)$ CPS [11-20].

Recently, the author of the present work [21] introduced a new class of CPSs called the conformable fractional $E^{\alpha}(5)$ CPS, where $\alpha$ is the order of the derivative in the conformable fractional BH. The analytic eigenvalue and eigenfunction solutions of conformable fractional BH (utilizing ISWP in $\beta$ variable) were obtained. The new $E^{\alpha}(5)$ CPS not only gave us the series of models with predictions directly comparable to the experiment (by modifying the value of $\alpha$), but also this approach clarified the way to converge from the classical $E(5)$ CPS. The present study aims to extend this type of results to the $U(5) - E(5)$ transition. We compare between the classical and conformable fractional solutions of BH using two potentials in this region. The low-lying energy spectra generated by the conformable fractional BH with free-parameter Bonatsos potentials, $\beta^{2n}/2$, and the two-parameters sextic potentials are investigated. The evolution of the probability density distributions, as a function of the collective variable $\beta$ and the order of the fractional derivative $\alpha$, is analyzed for the ground and excited states of these potentials. The Bonatsos potentials [22,23] provide a "bridge" between the classical $U(5)$ symmetry of the BH with a harmonic oscillator potential (HOP) and the $E(5)$ CPS of Iachello using BH with an ISWP. However, the conformable fractional spectra of these potentials close all the gaps between the classical spectra of the $U(5)$, $E(5) - \beta^{2n}$, and $E(5)$ models. The second potential considered in this study is the sextic potential in the $\beta$ variable [24-32]. It is characterized by, depending on the values of its parameters, a single spherical minimum, a deformed minimum, or simultaneous spherical and deformed minima separated by a barrier. By changing the barrier height, one can obtain the shape coexistence phenomena (the existence of states with different shapes in a narrow range of energies), the shape mixing phenomena, or the shape fluctuations phenomena (the wave function peaks begin to combine and merge but remain distinguishable), particularly for the CPSs of a QSPT when the height of the barrier is small and close to zero. The evolution of the conformable fractional spectra of the sextic potentials, as a function of barrier height and normalized to the energy of ground state, starts from the spectrum of the $O(6)$ dynamical symmetry to the spectrum of the $E(5)-\beta^6$ model. The conformable fractional BH with sextic potential is used for studying both the energy levels and the shape coexistence with mixing phenomena in the $^{96}$Mo nucleus. The results are in good agreement with the experimental findings.

The main point here is that the conformable fractional BH is able to generate many models similar to the original one, mainly when the value of $\alpha$ is very close to 1. The numerical results illustrated that these models are appropriate for analyzing a variety of phenomena at phase transitions. Consequently, we have a more explicit insight into the great variety of spectral and dynamical features at the QSPT. Clearly, such an approach is preferable because it allows expanding the original models in various ways and fits the experimental results with reasonable accuracy without developing a completely new model.

In section 2, the conformable fractional BH is introduced. In section 3, the evolution of the classical and conformable fractional spectra and wave functions of Bonatsos and sextic potential is given as a function of the fractional derivative order and potential parameters. Section 4 is devoted to discussing the energy spectrum in $^{96}$Mo nucleus, using the conformable fractional calculus. In section 5, our results are briefly summarized.

## 2. Conformable Fractional BH

Fractional calculus [33-35] is a well-defined branch of mathematics, which has been used in several areas of physics, including rotational and vibrational spectra in atomic nuclei. Many definitions of the fractional derivative, such as Caputo, Riemann-Liouville, Riesz, and Grunwald–Letnikov, have been proposed. In 2014, a new simple and captivating definition of fractional derivative had been suggested by Khalil et al. [36], namely conformable fractional derivative, that depends on the limit definition of the derivative. The advantage of using the conformable fractional calculus lies in the fact that several properties of the usual derivatives preserve their form within the framework of the conformable fractional calculus, while this is not the case within fractional calculus in general. These properties make the determination of analytical solutions for spectra and wave functions feasible. Many essential differential equations have been redefined due to the new conformable fractional derivative formulation [37-39]. The conformable fractional derivative of order $\alpha$ is defined for a function $f: [s, \infty) \to \mathbb{R}$ by

$$(D_s^\alpha f)(x) = \lim_{\varepsilon \to 0} \frac{f(x + \varepsilon(x-s)^{1-\alpha}) - f(x)}{\varepsilon}, \quad (1)$$

for all $x > s$, $\alpha \in (0,1]$. When $s = 0$, it is denoted by $D^\alpha$. The main properties and advantages of the conformable fractional derivative can be summarized as follow

(1) $D^\alpha(ch + dg) = cD^\alpha h + dD^\alpha g$, for all $c, d \in \mathbb{R}$ and $h, g$ be $\alpha$-differentiable at a point $x > 0$
(2) $D^\alpha x^t = tx^{t-\alpha}$, for all $t \in \mathbb{R}$
(3) $D^\alpha \rho = 0$, with $\rho$ is a constant
(4) $D^\alpha(hg) = hD^\alpha(g) + gD^\alpha(h)$
(5) $D^\alpha(h/g) = [gD^\alpha(h) - hD^\alpha(g)]/g^2$

The original BH will be our starting point:

$$H = -\frac{\hbar^2}{2B}\left[\frac{1}{\beta^4}\frac{\partial}{\partial \beta}\beta^4\frac{\partial}{\partial \beta} + \frac{1}{\beta^2 \sin 3\gamma}\frac{\partial}{\partial \gamma}\sin 3\gamma \frac{\partial}{\partial \gamma} - \frac{1}{4\beta^2}\sum_{\kappa=1,2,3}\frac{\hat{Q}_\kappa^2}{\sin^2(\gamma - 2\pi\kappa/3)}\right] + V(\beta, \gamma), \quad (2)$$

where $\beta, \gamma$ are the intrinsic variables, $B$ is the mass parameter, $\hat{Q}_k$ are the body-fixed components of the angular momentum, and $\hbar$ is the reduced Planck constant. Following the standard procedure, for a $\gamma$-independent potential, i.e., $V(\beta,\gamma) = U(\beta)$, one can separate the variables by assuming

$$\Psi(\beta, \gamma, \theta_i) = f(\beta)\Phi(\gamma, \theta_i), \quad (3)$$

where $\theta_i$ ($i = 1,2,3$) are the three Euler angles and the functions $f(\beta)$ depend on the choice of the $U(\beta)$ potential. Now, we can separate the Hamiltonian (2) into two equations:

$$\left[-\frac{1}{\sin 3\gamma}\frac{\partial}{\partial \gamma}\sin 3\gamma \frac{\partial}{\partial \gamma} + \frac{1}{4}\sum_\kappa \frac{\hat{Q}_\kappa^2}{\sin^2(\gamma - 2\pi\kappa/3)}\right]\Phi(\gamma, \theta_i) = \Lambda\Phi(\gamma, \theta_i), \quad \Lambda = \tau(\tau + 3), \tau = 0,1,2,... \quad (4.1)$$

$$\left[-\frac{\hbar^2}{2B}\left(\frac{1}{\beta^4}\frac{\partial}{\partial\beta}\beta^4\frac{\partial}{\partial\beta} - \frac{\Lambda}{\beta^2}\right) + U(\beta)\right]f(\beta) = Ef(\beta), \tag{4.2}$$

Let us introduce the reduced energies $\varepsilon = \frac{2B}{\hbar^2}E$ and reduced potentials $u = \frac{2B}{\hbar^2}U$. The radial equation in the $\beta$ variable, (4.2), becomes

$$\left[-\frac{1}{\beta^4}\frac{\partial}{\partial\beta}\beta^4\frac{\partial}{\partial\beta} + \frac{\Lambda}{\beta^2} + u(\beta)\right]f(\beta) = \varepsilon f(\beta). \tag{5}$$

Under the assumptions that $\varphi(\beta) = \beta^{3/2}f(\beta)$, we have

$$\varphi''(\beta) + \frac{\varphi'(\beta)}{\beta} + \left\{\varepsilon - u(\beta) - \frac{(\tau + 3/2)^2}{\beta^2}\right\}\varphi(\beta) = 0. \tag{6}$$

The values of angular momentum $L$ included in each value of $\tau$, (i.e., included in each irreducible representation of $SO(5)$), are given as follows: $\tau = 3\nu_\Delta + \lambda$, where $\nu_\Delta = 0,1,\ldots$ is the missing label in the reduction $SO(5) \supset SO(3)$, and taking $L = \lambda, \lambda + 1, \ldots, 2\lambda - 2, 2\lambda$ (note that $2\lambda - 1$ missing).

Equation (6) can be rewritten in conformable fractional form by replacing the orders of differentiations and the polynomial coefficients by their non-integer order. This operation is based on a two-stage strategy; deduce a valid formula for the integer value $I, I \in \mathbb{N}$, then replace $I$ with fractional-order $\alpha$. Using this strategy, (6) becomes

$$D^\alpha D^\alpha \varphi(\beta) + \frac{1}{\beta^\alpha}D^\alpha \varphi(\beta) + \left\{\varepsilon - u(\beta) - \frac{(\tau + 3/2)^2}{\beta^{2\alpha}}\right\}\varphi(\beta) = 0. \tag{7}$$

Using the following properties of conformable fractional derivative,

$$D^\alpha \varphi(\beta) = \beta^{1-\alpha}\frac{d\varphi(\beta)}{d\beta}, \qquad D^\alpha D^\alpha \varphi(\beta) = (1-\alpha)\beta^{1-2\alpha}\frac{d\varphi(\beta)}{d\beta} + \beta^{2-2\alpha}\frac{d^2\varphi(\beta)}{d\beta^2}, \tag{8}$$

and replacing these definitions in (7), the general conformable fractional BH in collective variable $\beta$ is obtained;

$$\frac{d^2\varphi(\beta)}{d\beta^2} + \frac{(2-\alpha)}{\beta}\frac{d\varphi(\beta)}{d\beta} + \frac{1}{\beta^{2-2\alpha}}\left\{\varepsilon - u(\beta) - \frac{(\tau + 3/2)^2}{\beta^{2\alpha}}\right\}\varphi(\beta) = 0. \tag{9}$$

Among the various approaches used for solving differential equations, numerical methods that use grid-based strategies became standard methods. These methods produce excellent results with a relatively low computational effort. The differential equation can be discretized using a mesh with well-spaced grid points. In finite difference methodology, the derivatives in the differential equation are approximated using finite difference formulas. We refer the reader to the references [40-44] for further details on the finite difference method. The solution domain $[a, b]$ is divided into $N$ equal-length subintervals, which have been described by $(N + 1)$ grid points. At each interior grid point of the domain, the differential equation is written. Consequently, a system of algebraic equations arises. The numerical solution of the differential equation is actually the solution of that system. The central difference formulas are often used because they have high accuracy. The first and the second derivatives at the interior grid points are defined by:

$$\frac{d\varphi}{d\beta} = \frac{\varphi_{i+1} - \varphi_{i-1}}{2h}, \qquad \frac{d^2\varphi}{d\beta^2} = \frac{\varphi_{i-1} - 2\varphi_i + \varphi_{i+1}}{h^2}. \tag{10}$$

where $h$ is the step size. The finite-difference representation for (9) becomes:

$$\frac{\varphi_{i-1} - 2\varphi_i + \varphi_{i+1}}{h^2} + \frac{(2-\alpha)}{\beta_i}\frac{\varphi_{i+1} - \varphi_{i-1}}{2h} + \frac{1}{\beta_i^{2-2\alpha}}\left\{\varepsilon - u(\beta_i) - \frac{(\tau + 3/2)^2}{\beta_i^{2\alpha}}\right\}\varphi_i = 0. \tag{11}$$

This equation creates a system of $N - 1$ algebraic equations for the unknowns $\varphi_2, \ldots, \varphi_N$. The implicit Dirichlet boundary conditions are $\varphi_1 = \varphi_{N+1} = 0$. Equation (11) is a simplified eigenvalue problem that can be solved for the lowest eigenfunctions and energy eigenvalues. The Gnu Scientific Library, ARPACK library, and MKL are only few examples of frameworks that provide effective algorithms.

## 3. Bonatsos and sextic potentials with conformable fractional BH

In this section, we investigate two classes of potentials that play an essential role in nuclear structure. The first class of potentials is the Bonatsos potentials of the form

$$H_n(\beta) = \beta^{2n}/2, \tag{12}$$

where $n = 2,3,4$. For $n = 1$, we get the HOP whilst the ISWP manifests for $n \to \infty$. The finite difference method is used to calculate the conformable fractional spectra of the $H_n(\beta)$ potentials. The second-order differential equation, (9), is solved for each value of $\tau = 0,1,2, \ldots$ separately. For a specific value of the order of the fractional derivative, $\alpha$, the energy levels are labeled by $(L, \xi, \tau)$ quantum numbers, where $\xi$ labels the main families of energy levels ($\xi = 1,2,3,...$), and $\tau$ characterizes the phonon-like energy levels inside $\xi$ family ($\tau = 0,1,2,...$). For instance, the $2^+_1$, $4^+_1$, and $2^+_2$ levels are denoted by $2^+_{1,1}$, $4^+_{1,2}$, and $2^+_{1,2}$, respectively. The energy spectrum is normalized to the $2^+_{1,1}$ state of the ground band as the following,

$$\epsilon^\alpha_{L^+_{\xi,\tau}} = \frac{E^\alpha_{L^+_{\xi,\tau}} - E^\alpha_{0^+_{1,0}}}{E^\alpha_{2^+_{1,1}} - E^\alpha_{0^+_{1,0}}}. \tag{13}$$

**Table 1.** Spectra of the fractional $E^\alpha(5)$-$\beta^4$ models, $\alpha = 0.5, 0.7$, and $0.9$, compared to the predictions of the classical $U(5)$ symmetry and Bonatsos $E(5)$-$\beta^4$ models.

| $L^+_{\xi,\tau}$ | | | | $U(5)$ | $E^{\alpha=0.5}(5)$-$\beta^4$ | $E^{\alpha=0.7}(5)$-$\beta^4$ | $E^{\alpha=0.9}(5)$-$\beta^4$ | $E(5)$-$\beta^4$ ($\alpha=1$) |
|---|---|---|---|---|---|---|---|---|
| $0^+_{1,0}$ | | | | 0.00000 | 0.00000 | 0.00000 | 0.00000 | 0.00000 |
| $2^+_{1,1}$ | | | | 1.00000 | 1.00000 | 1.00000 | 1.00000 | 1.00000 |
| $4^+_{1,2}$ | $2^+_{1,2}$ | | | 2.00000 | 2.06084 | 2.07431 | 2.08676 | 2.09274 |
| $6^+_{1,3}$ | $4^+_{1,3}$ | $3^+_{1,3}$ | $0^+_{1,3}$ | 3.00000 | 3.17009 | 3.21036 | 3.24737 | 3.26497 |
| $0^+_{2,0}$ | | | | 2.00000 | 2.23910 | 2.30034 | 2.35993 | 2.38957 |
| $2^+_{2,1}$ | | | | 3.00000 | 3.36844 | 3.47445 | 3.57572 | 3.62492 |
| $4^+_{2,2}$ | $2^+_{2,2}$ | | | 4.00000 | 4.53685 | 4.69409 | 4.84486 | 4.91783 |
| $6^+_{2,3}$ | $4^+_{2,3}$ | $3^+_{2,3}$ | $0^+_{2,3}$ | 5.00000 | 5.74031 | 5.95595 | 6.16491 | 6.26594 |
| $0^+_{3,0}$ | | | | 4.00000 | 4.68247 | 4.87402 | 5.06129 | 5.15324 |
| $2^+_{3,1}$ | | | | 5.00000 | 5.91089 | 6.17568 | 6.43749 | 6.56475 |
| $4^+_{3,2}$ | $2^+_{3,2}$ | | | 6.00000 | 7.17547 | 7.50923 | 7.85085 | 8.01683 |
| $6^+_{3,3}$ | $4^+_{3,3}$ | $3^+_{3,3}$ | $0^+_{3,3}$ | 7.00000 | 8.47982 | 8.87478 | 9.30248 | 9.51075 |
| $0^+_{4,0}$ | | | | 6.00000 | 7.30921 | 7.66108 | 8.03324 | 8.21601 |
| $2^+_{4,1}$ | | | | 7.00000 | 8.65099 | 9.06627 | 9.53920 | 9.77088 |
| $4^+_{4,2}$ | $2^+_{4,2}$ | | | 8.00000 | 10.0490 | 10.4975 | 11.0727 | 11.3562 |
| $6^+_{4,3}$ | $4^+_{4,3}$ | $3^+_{4,3}$ | $0^+_{4,3}$ | 9.00000 | 11.5118 | 11.9587 | 12.6369 | 12.9752 |

**Table 2.** Spectra of the fractional $E^\alpha(5)$-$\beta^6$ models, $\alpha = 0.5, 0.7$, and $0.9$, compared to the predictions of the classical Bonatsos $E(5)$-$\beta^4$ and $E(5)$-$\beta^6$ models.

| $L^+_{\xi,\tau}$ | | | | $E(5)$-$\beta^4$ | $E^{\alpha=0.5}(5)$-$\beta^6$ | $E^{\alpha=0.7}(5)$-$\beta^6$ | $E^{\alpha=0.9}(5)$-$\beta^6$ | $E(5)$-$\beta^6$ ($\alpha=1$) |
|---|---|---|---|---|---|---|---|---|
| $0^+_{1,0}$ | | | | 0.00000 | 0.00000 | 0.00000 | 0.00000 | 0.00000 |
| $2^+_{1,1}$ | | | | 1.00000 | 1.00000 | 1.00000 | 1.00000 | 1.00000 |
| $4^+_{1,2}$ | $2^+_{1,2}$ | | | 2.09274 | 2.12134 | 2.12638 | 2.13180 | 2.13470 |
| $6^+_{1,3}$ | $4^+_{1,3}$ | $3^+_{1,3}$ | $0^+_{1,3}$ | 3.26497 | 3.34767 | 3.36423 | 3.38129 | 3.39013 |
| $0^+_{2,0}$ | | | | 2.38957 | 2.54115 | 2.56996 | 2.60154 | 2.61871 |
| $2^+_{2,1}$ | | | | 3.62492 | 3.86244 | 3.92244 | 3.98246 | 4.01284 |
| $4^+_{2,2}$ | $2^+_{2,2}$ | | | 4.91783 | 5.26609 | 5.36024 | 5.45311 | 5.49949 |
| $6^+_{2,3}$ | $4^+_{2,3}$ | $3^+_{2,3}$ | $0^+_{2,3}$ | 6.26594 | 6.74624 | 6.87975 | 7.01090 | 7.07615 |
| $0^+_{3,0}$ | | | | 5.15324 | 5.60604 | 5.71684 | 5.83068 | 5.88951 |
| $2^+_{3,1}$ | | | | 6.56475 | 7.17598 | 7.34447 | 7.51127 | 7.59484 |
| $4^+_{3,2}$ | $2^+_{3,2}$ | | | 8.01683 | 8.80948 | 9.03739 | 9.26105 | 9.37219 |
| $6^+_{3,3}$ | $4^+_{3,3}$ | $3^+_{3,3}$ | $0^+_{3,3}$ | 9.51075 | 10.5041 | 10.7961 | 11.0818 | 11.2233 |
| $0^+_{4,0}$ | | | | 8.21601 | 9.10492 | 9.34709 | 9.59107 | 9.71499 |
| $2^+_{4,1}$ | | | | 9.77088 | 10.8840 | 11.2081 | 11.5279 | 11.6874 |
| $4^+_{4,2}$ | $2^+_{4,2}$ | | | 11.3562 | 12.7147 | 13.1217 | 13.5207 | 13.7188 |
| $6^+_{4,3}$ | $4^+_{4,3}$ | $3^+_{4,3}$ | $0^+_{4,3}$ | 12.9752 | 14.5968 | 15.0907 | 15.5741 | 15.8135 |

The predictions for the conformable fractional spectra normalized according to (13) for the potentials $H^{\alpha=1}_1 \equiv$ HOP, $H^\alpha_2$, $H^\alpha_3$, $H^\alpha_4$, and $H^{\alpha=1}_\infty \equiv E(5)$ are shown in tables 1-3. This sequence of potentials serves as a bridge between the HOP with $U(5)$ symmetry and the ISWP with $E(5)$ symmetry. The classical energy spectra ($\alpha = 1$) do not include any free parameters giving valuable benchmarks to compare with other models. Tables 1-3 also show that as $\alpha$ increases, the spectra approach the classical $E(5) - \beta^{2n}$ models smoothly, across all bands and for all values of the angular momentum, $L$. The same result is shown in figure 1(a), where multiple energy levels of the ground state band within each model (with $\alpha = 0.5$ and 1) are presented as a function of $L$, as well as from figure 1(b), which shows the band-heads of various excited bands for each model against the index $\xi$.

**Table 3.** Spectra of the fractional $E^\alpha(5)$-$\beta^6$ models, $\alpha = 0.5, 0.7$, and $0.9$, compared to the predictions of the classical Bonatsos $E(5)$-$\beta^6$ and $E(5)$-$\beta^8$ models.

| $L^+_{\xi,\tau}$ | | | | $E(5)$-$\beta^6$ | $E^{\alpha=0.5}(5)$-$\beta^8$ | $E^{\alpha=0.7}(5)$-$\beta^8$ | $E^{\alpha=0.9}(5)$-$\beta^8$ | $E(5)$ ($\alpha=1$) |
|---|---|---|---|---|---|---|---|---|
| $0^+_{1,0}$ | | | | 0.00000 | 0.00000 | 0.00000 | 0.00000 | 0.00000 |
| $2^+_{1,1}$ | | | | 1.00000 | 1.00000 | 1.00000 | 1.00000 | 1.00000 |
| $4^+_{1,2}$ | $2^+_{1,2}$ | | | 2.13470 | 2.15119 | 2.15280 | 2.15526 | 2.19857 |
| $6^+_{1,3}$ | $4^+_{1,3}$ | $3^+_{1,3}$ | $0^+_{1,3}$ | 3.39013 | 3.43815 | 3.44459 | 3.45290 | 3.58978 |
| $0^+_{2,0}$ | | | | 2.61871 | 2.71947 | 2.72961 | 2.74536 | 3.03126 |
| $2^+_{2,1}$ | | | | 4.01284 | 4.17105 | 4.20287 | 4.23807 | 4.80009 |
| $4^+_{2,2}$ | $2^+_{2,2}$ | | | 5.49949 | 5.73462 | 5.78900 | 5.84640 | 6.78048 |
| $6^+_{2,3}$ | $4^+_{2,3}$ | $3^+_{2,3}$ | $0^+_{2,3}$ | 7.07615 | 7.40374 | 7.48391 | 7.56727 | 8.96753 |
| $0^+_{3,0}$ | | | | 5.88951 | 6.20240 | 6.26232 | 6.33205 | 7.57750 |
| $2^+_{3,1}$ | | | | 7.59484 | 8.01890 | 8.12031 | 8.22761 | 10.1083 |
| $4^+_{3,2}$ | $2^+_{3,2}$ | | | 9.37219 | 9.92797 | 10.0709 | 10.2184 | 12.8556 |
| $6^+_{3,3}$ | $4^+_{3,3}$ | $3^+_{3,3}$ | $0^+_{3,3}$ | 11.2233 | 11.9267 | 12.1143 | 12.3060 | 15.8161 |
| $0^+_{4,0}$ | | | | 9.71499 | 10.3473 | 10.4965 | 10.6588 | 13.6407 |
| $2^+_{4,1}$ | | | | 11.6874 | 12.4850 | 12.6947 | 12.9123 | 16.9330 |
| $4^+_{4,2}$ | $2^+_{4,2}$ | | | 13.7188 | 14.7024 | 14.9724 | 15.2481 | 20.4438 |
| $6^+_{4,3}$ | $4^+_{4,3}$ | $3^+_{4,3}$ | $0^+_{4,3}$ | 15.8135 | 16.9980 | 17.3313 | 17.6694 | 24.1718 |

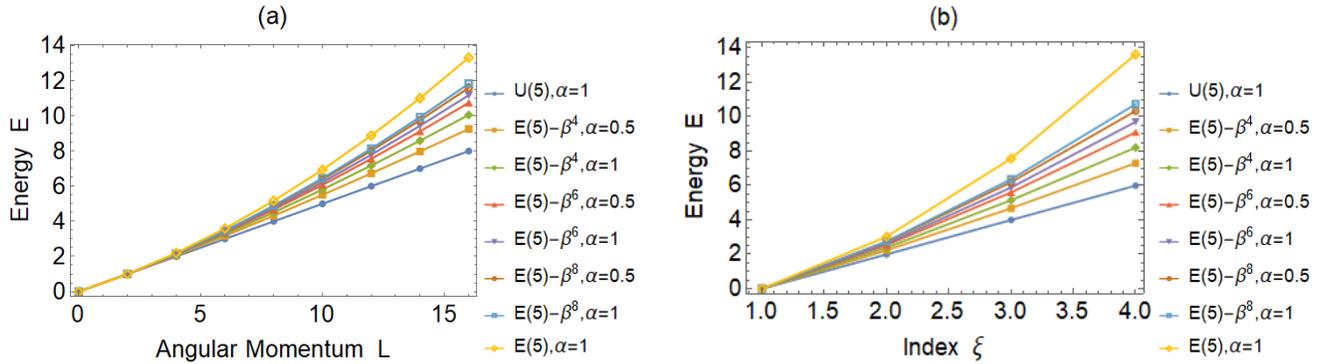

**Figure 1.** (Color online) (a) The energy levels of the ground state band of the $U(5)$, $E^\alpha(5)$-$\beta^4$, $E^\alpha(5)$-$\beta^6$, $E^\alpha(5)$-$\beta^8$ and $E(5)$ models (with $\alpha = 0.5$ and 1) as a function of the angular momentum $L$. The energy spectrum is normalized to the energy of the $2^+_{1,1}$ state (13). (b) Band-head energies of excited bands for the same models against the index $\xi$.

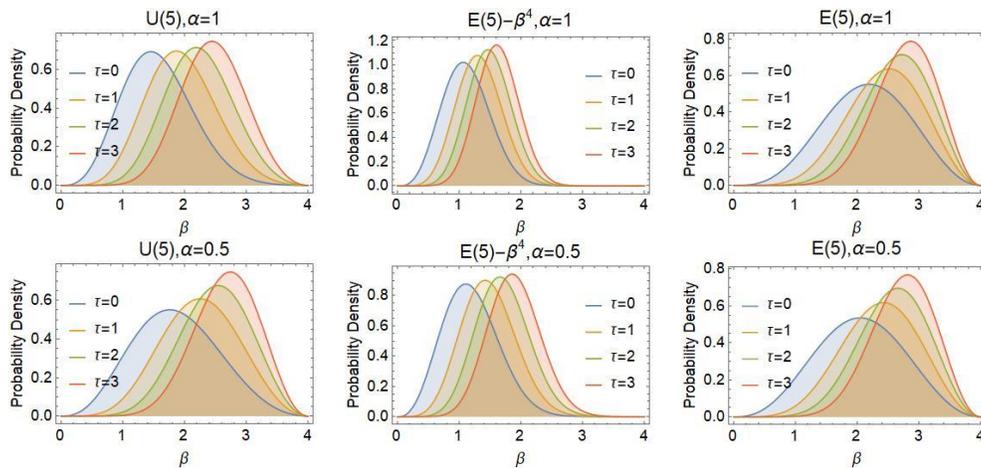

**Figure 2.** (Color online) The evolution of the probability density distributions of the $\tau = 0, 1, 2, 3$ ground band states as a function of the collective variable $\beta$, as well as the order of the fractional derivative $\alpha$, of the $U(5)$, $E^\alpha(5)$-$\beta^4$, and $E(5)$ models.

**Table 4.** Some numerical values of the parameters $a$ and $b$ of the sextic potential generate the $\tau = 0$ effective potential with two degenerate minima.

| $a$ | 0.02 | 0.03 | 0.04 | 0.05 | 0.06 | 0.07 | 0.08 | 0.09 | 0.10 | 0.11 |
|---|---|---|---|---|---|---|---|---|---|---|
| $b$ | 0.000103 | 0.000236 | 0.000427 | 0.00068 | 0.000998 | 0.00138 | 0.001835 | 0.00237 | 0.00299 | 0.0037 |

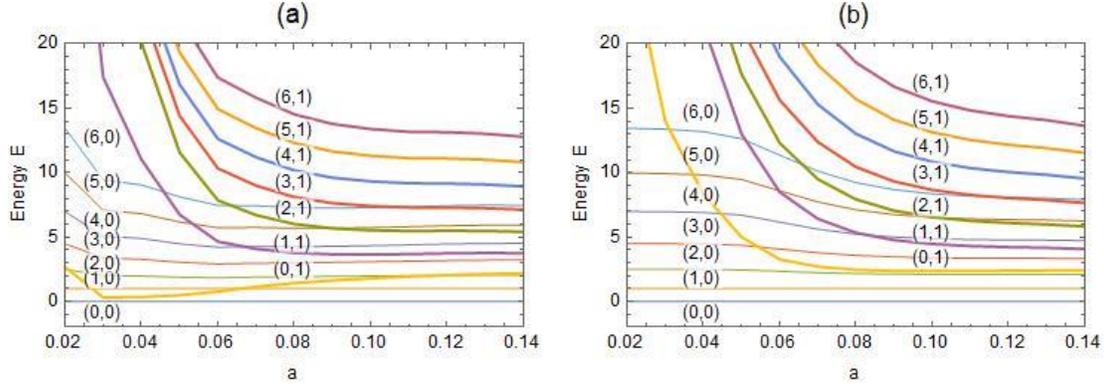

**Figure 3.** (Color online) (a) The conformable fractional (with $\alpha = 1$, or classical) energy levels of the ground state band and first excited $\beta$ band of the sextic potential against the values of the parameter $a$ limited by the constraints of table 4. The energy levels are labeled by $(\tau, \xi)$. (b) The conformable fractional spectrum with $\alpha = 0.9$. The spectra normalized according to (13).

The essential signature of the classical $E(5)$ CPS is the energy ratio $R(4/2) = \epsilon_{4^+_{1,2}}/\epsilon_{2^+_{1,1}} = 2.20$; it is in the middle between $(R(4/2) = 2.00)$ of the classical spherical vibrator and $(R(4/2) = 2.50)$ of the classical deformed $\gamma$-unstable. The classical energy ratio of the $\beta^4/2$, $\beta^6/2$, and $\beta^8/2$ potentials, with $\alpha = 1$, are $R(4/2) = 2.09274, 2.13470$, and $2.15526$, respectively. However, the conformable fractional energy ratio of these potentials, with $\alpha$ in the range 0.5-0.9, are 2.06084-2.08676, 2.12134-2.13180, 2.15119-2.15526, respectively. Hence, the conformable fractional spectra of these potentials close all the gaps between the classical spectra of the $U(5)$, $E(5) - \beta^{2n}$, and $E(5)$ models. The same result is confirmed in figure 2, where the evolution of the probability density distributions, as a function of the collective variable $\beta$ and the order of the fractional derivative $\alpha$, is presented for the ground and excited states of the potentials $H_1^\alpha \equiv$ HOP, $H_2^\alpha$, and $H_\infty^\alpha \equiv E(5)$.

It is important to note that, $H_n(\beta)$ has only one minimum. For the transition between two quantum phases, such potentials are ineffective, particularly when one of the phases due to HOP. In this case, the critical potential for a phase transition is an entirely flat profile, i.e., ISWP. A fundamental question must be asked: are there any other more flexible potentials that can effectively reproduce the characteristics of the ISWP which are dependent on a small number of parameters while still presenting new features? Good candidates are sextic potentials. Such potentials are of the form

$$V(\beta) = \beta^2 - a\beta^4 + b\beta^6, \tag{14}$$

where $a$ and $b$ are two parameters, the effective potential in this case becomes:

$$v_{eff}(\beta) = \frac{1}{\beta^{2-2\alpha}}\left(\beta^2 - a\beta^4 + b\beta^6 + \frac{(\tau + 3/2)^2}{\beta^{2\alpha}}\right). \tag{15}$$

It is challenging to create a complete map of model characteristics of a sextic potential as a function of both $a$ and $b$. However, one can pick an interesting path in the parameter space and analyze the evolution of the model characteristics along this path. Here we are interested in what occurs when the effective potential with $\tau = 0$, which generates both the ground state and the $\beta$ excited states $0^+$, has two degenerated minima. This constraint establishes a numerical relationship between the parameters $a$ and $b$, which is listed in table 4. The evolution of the low-lying energy spectrum normalized according to (13) of the ground band and first excited $\beta$ band as a function of parameter $a$, constrained by the values of table 4, is depicted in figure 3. From there, it can be seen that the energy levels from the two bands have become almost degenerate in the second half of the interval under consideration. Also, as $a$ changes in this interval, the states from the two bands alternate their order and intersect each other at some point. The energy spectrum in figure 3 illustrates a transformation from an $O(6)$ energy level sequence to the spectrum of $E(5) - \beta^6$ model. Such limiting models can efficiently be recovered by using reasonable limits for the parameters of the sextic potential (15). The $O(6)$ limit is approached when the parameter $a$ has small values. In this case the sexic potential forms a narrow deformed minimum separated by a high barrier. The $E(5) - \beta^6$ model is approached for high values of $a$, which makes the $\beta^6$ term dominant.

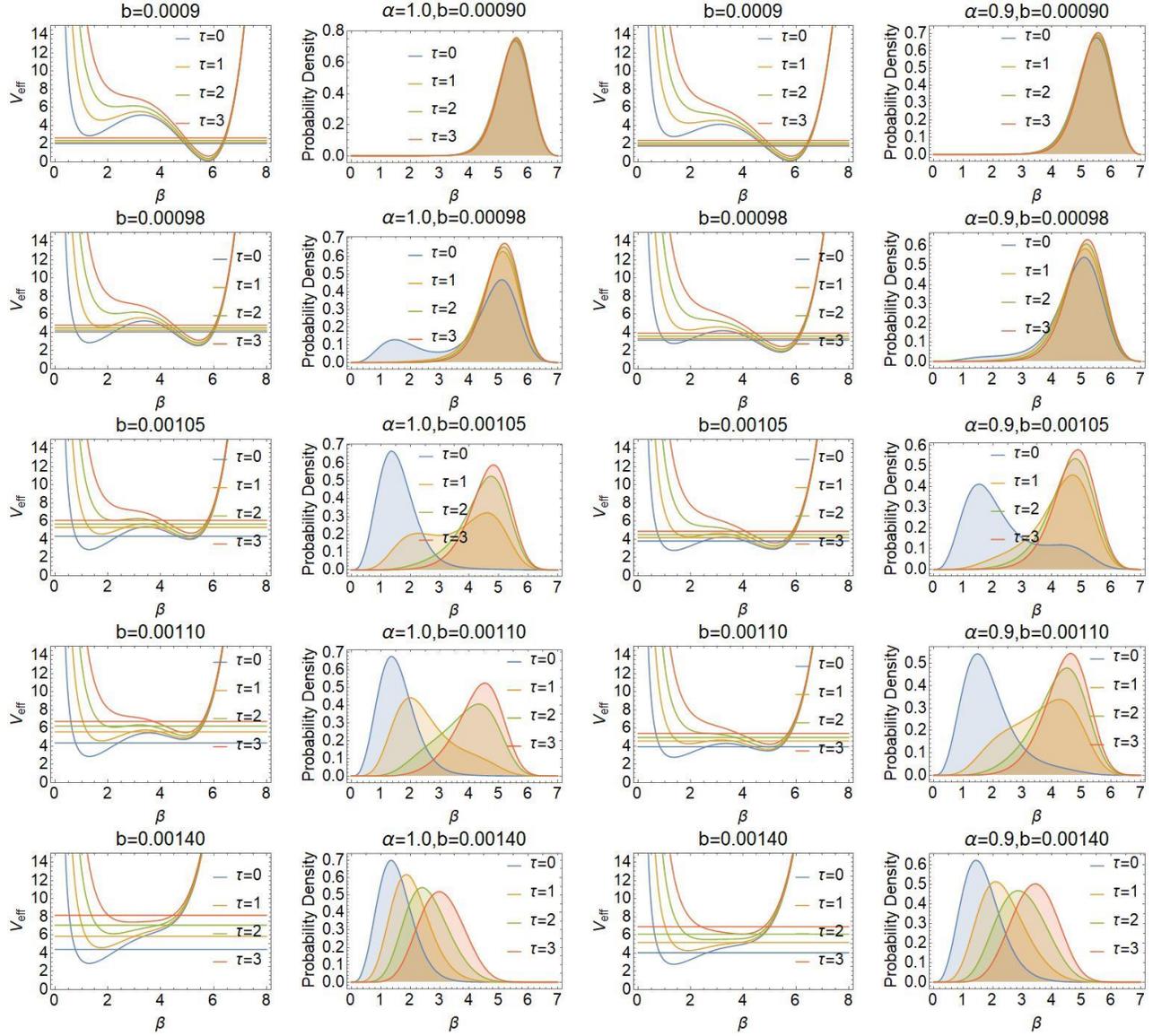

**Figure 4.** (Color online) Matrix of effective potentials of the sextic potential with $\alpha = 1.0$ (first column), probability density functions corresponding to the $\tau = 0,1,2,3$ ground band states as a function of the deformation variable $\beta$, computed using sextic potential with $a = 0.06$ and $b = 0.0090$-$0.00140$, for $\alpha = 1.0$ (second column), effective potentials with $\alpha = 0.9$ (third column), and probability density functions with $\alpha = 0.9$ (last column).

The probability density distribution can be used to get a deeper understanding of these models. Figure 4 shows the development of the effective potentials, classical ($\alpha = 1$) probability density distributions, and the conformable fractional ($\alpha = 0.9$) probability distributions for the ground band states with $\xi = 0$ and $\tau = 0,1,2,3$ of the sextic potentials (15), as a function of the collective variable $\beta$ and the parameter $b$, in the case of $a = 0.06$. At $b = 0.0009$, the classical probability density distributions of $\tau = 0,1,2,3$ states become narrow and mainly centered around the deeper minimum of each potential. For $b = 0.00098$, due to the tunneling between the potential wells, the two peaks (with unequal height) of the probability density of $\tau = 0$ state become closely spaced. However, the energy levels with $\tau = 1,2,3$ are still below the second minimum of the corresponding effective potential. The probability density functions of these states are also centered on the deeper minimum of each potential. For $b = 0.00105$, the probability density functions of the $\tau = 0$ state extend over two minima and have a high peak around one of them, but its tail extends to the other minimum, the two peaks (with unequal height) of the probability density of $\tau = 1$ state become closely spaced, and in the same time, the probability density functions of the $\tau = 2,3$ states are centered on the deeper minimum of the corresponding potential. Decreasing further the barrier, at $b = 0.00110$, the probability density functions of the $\tau = 1,2$ states extend over two minima and have a high peak around one of them, but its tail extends to the other minimum. However, the probability density functions of the $\tau = 0,3$ states are centered on the deeper minimum of the corresponding potential. At $b = 0.00140$, the $\tau = 0,1,2,3$ states are above the barrier, each potential has approximately one minimum (the barrier becomes very weak), and one obtains a probability distribution with a single smooth peak. Actually, as seen in this picture, the shape mixing phenomena, the shape coexistence scenarios, and the fluctuations phenomena can be generated by changing the height of the barrier. An important example is the coexistence of shapes consistent with the low deformation of the $\tau = 2,3$ states and the high deformation of the $\tau = 0,1$ state of the $\xi = 0$ band at $\alpha = 1$, $b = 0.00105$. We emphasize that our aim, in this section, is not to reproduce the experimental results but rather to obtain a qualitative image of the general performance of the classical and conformable fractional models. Experimental evidence will be discussed in the following section.

**Table 5.** Low energy theoretical and experimental spectra of $^{96}$Mo [45] normalized to the energy of the $2^+_{1,1}$ state, see (13). The theoretical values are the results from (9) and (15) for $a = 0.036$ and $b = 0.0004$, $\alpha = 0.965$ and $\alpha = 1$.

| | $^{96}$Mo | | |
|---|---|---|---|
| $L^+_{\xi,\tau}$ | $\epsilon^{\exp}_{L^+_{\xi,\tau}}$ | $\epsilon^{\alpha=0.965}_{L^+_{\xi,\tau}}$ | $\epsilon^{\alpha=1}_{L^+_{\xi,\tau}}$ |
| $0^+_{1,0}$ | 0.00000 | 0.00000 | 0.00000 |
| $2^+_{1,1}$ | 1.00000 | 1.00000 | 1.00000 |
| $4^+_{1,2}$ | 2.09215 | 1.90081 | 1.92556 |
| $2^+_{1,2}$ | 2.08922 | 1.90081 | 1.92556 |
| $6^+_{1,3}$ | 3.13627 | 2.49066 | 2.71991 |
| $4^+_{1,3}$ | 2.40232 | 2.49066 | 2.71991 |
| $3^+_{1,3}$ | 2.54222 | 2.49066 | 2.71991 |
| $0^+_{2,0}$ | 1.47513 | 1.71317 | 1.77427 |
| $2^+_{2,1}$ | 1.92459 | 2.19215 | 2.45020 |
| Δ | | 0.430872 | 0.479193 |
| σ | | 0.264078 | 0.284033 |

We can obviously observe that the probability distribution and energy spectrum are directly affected by the height of the barrier. The introduction of the barrier significantly changes the density probability distribution. For instance, in the case of the ISWP, the probability density distribution for the ground state has a symmetric form with a large width, and the peak is located near the middle of the potential. However, when the barrier is introduced and increased, the peaks become very narrow and more centered around the minima, as seen in figure 4.

The values of the energy levels generated by the conformable fractional ($\alpha = 0.9$) BH are lower than the values of the corresponding classical energy levels ($\alpha = 1$), see figure 4. Similarly, this occurs with the effective potentials. The behavior of the conformable fractional probability distributions for the ground band states with $\tau = 0,1,2,3$ and energy spectra are quite close and similar to the classical one, although the energy levels are shifted down. The main idea here is that the conformable fractional BH contains a free parameter which is the order of the conformable fractional derivative. Hence, we have many models similar to the original one (particularly when the value of $\alpha$ is very close to 1), this increases the possibility that one of them represents the experimental results with satisfactory accuracy.

## 4. Experimental realization

It is essential to see if the conformable fractional BH predictions are supported by experimental evidence. Table 5 compares experimental data for the different energy levels of the $^{96}$Mo nucleus to theoretical predictions. The root mean absolute error, Δ, and root-mean-square deviation, σ, are used to determine the fit quality. They are defined by

$$\Delta = \sqrt{\sum_i \left|\epsilon^{\exp}_{L^+_{\xi,\tau}} - \epsilon^{\alpha}_{L^+_{\xi,\tau}}\right|/n_l}, \quad \sigma = \sqrt{\sum_i \left(\epsilon^{\exp}_{L^+_{\xi,\tau}} - \epsilon^{\alpha}_{L^+_{\xi,\tau}}\right)^2/n_l}, \tag{16}$$

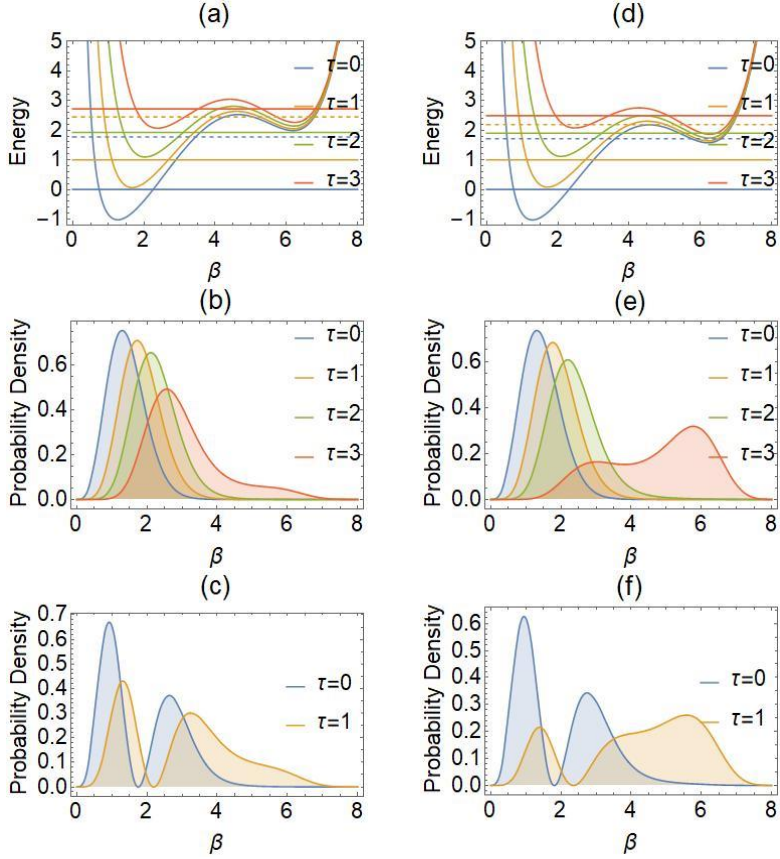

**Figure 5.** (Color online) Matrix of theoretical effective potentials ((a) and (d)), scaled to the energy of the $2^+_{1,1}$ state, probability density functions corresponding to the $\tau = 0,1,2,3$ states of the ground band ((b) and (e)) and $\beta$ excited states with $\tau = 0, 1$ ((c) and (f)) as a function of the deformation variable $\beta$, computed using sextic potential with $a = 0.036$ and $b = 0.0004$, for $\alpha = 1.0$ (first column) and $\alpha = 0.965$ (second column) for the $^{96}$Mo nucleus. The energy levels for $\xi = 0$ and $\xi = 1$ states are presented with solid and dashed lines, respectively.

where $n_l$ denotes the number of fitted energy levels while $\epsilon^{\text{exp}}_{L^+_{\xi,\tau}}$ and $\epsilon^{\alpha}_{L^+_{\xi,\tau}}$ represent the experimental and theoretical energies of the $i$th level. Once the diagonalizations are achieved for all angular momenta, a spectrum is obtained, and the functions $\Delta$ and $\sigma$ can be defined. The model parameters $a$, $b$ and $\alpha$ can then be modified to provide the most accurate representation of the experimental data. Since the finite-difference technique is dependent on the boundary $\beta_b$, it must be optimized beforehand for each set of the potential parameters. The results of the conformable fractional BH with sextic potentials reproduce the structure of the energy levels in the $^{96}$Mo nucleus and suggest more exact outcomes, i.e., minimum of the $\Delta$ and $\sigma$ in fractional BH, in comparison with the corresponding values in classical BH.

The structure and dynamics of quadrupole deformation associated with each state can be obtained from a study of the corresponding effective potential (15) for that state. Figures 5(a) and 5(d) describe the low-lying spectrum of the $^{96}$Mo nucleus, using classical and conformable fractional BH with the corresponding effective sextic potentials. It is clear that the double-well nature of the effective potential remains for various values of $\tau$ on which the real angular momentum states are constructed. The $\tau = 0,1,2$ states of $\xi = 0$ and the $\beta$ excited state with $\tau = 0$ are almost entirely trapped by the deeper minimum of the corresponding effective potentials. The $\tau = 3$ state extends over two minima while remaining below the separating barrier. Because of the relatively small barrier and quantum tunneling, the wave function will exist even in the less deformed minimum. The case is better described when looking at the shapes of the probability distribution in figures 5(b) and 5(e). From figure 5(b), it is obvious that the classical probability density distributions of $\xi = 0$ and $\tau = 0,1,2$ states become narrow and centered around the deeper minimum of the corresponding potential. While the $\tau = 3$ state extends over two minima and has a high peak around one of them, but its tail extends to the other. From figure 5(e), we can see that the conformable fractional probability density distributions of $\xi = 0$ and

$\tau=0,1,2$ states are similar to the corresponding classical cases; however, the $\tau=3$ state extends over two minima and the two peaks (with unequal high) of the probability density become closely spaced. We have a very clear double-peak structure in the probability distribution with conformable fractional BH. Figures 5(c) and 5(f) show the classical and conformable fractional probability distributions, respectively, for $\xi=1$ and $\tau=0,1$ states of the $\beta$ excited band. In figures 5(c), for $\tau=0$, the $\beta$ excitation proceeds as a regular vibration restricted to the deeper minimum, and the probability distribution consists of smooth two parts. However, for $\tau=1$ $\beta$ excited state, the probability distribution is divided into two parts. The second part becomes more extended over two minima and has a peak around the one minimum; however, its tail spreads to the other one. It has a three-peak structure due to quantum tunneling. In figure 5(f), the conformable fractional probability density distribution of $\xi=1$ and $\tau=1$ state extends over two minima, and the three peaks (with unequal high) of the probability density become closely spaced.

This image demonstrates many coexisting phenomena. The first scenario is the mixing of shapes in the $\xi=0$, $\tau=3$ state. The second one is the coexistence of shapes consistent with the low deformation of the $\tau=0,1,2$ states and the high deformation of the $\tau=3$ state of the $\xi=0$ band. A comparison of the sextic potential results with those for HOP and ISWP confirms this conclusion. The $\xi=0$ and $\tau=0,1,2$ states of the sextic potential behave identically to the states of HOP, while the $\xi=0$ and $\tau=3$ state appears similar to the ISWP, see figure 2. In the $\beta$ excited band with $\xi=1$, the comparison holds only for the $\tau=0$ state, while the $\tau=1$ state is strongly affected by shape mixing.

## 5. Conclusions

A serious problem confronted in the study of QSPTs and CPSs in nuclei is that there is no continuous variable across the nuclear chart. Along a chain of even-even isotopes, in which a QSPT is expected to be seen, one jumps from one isotope to the next by adding two neutrons each time. If the QSPT is of the first order, like the $X(5)$ CPS from vibrational to rotational nuclei, the changes occurring at the critical point are sizeable; thus, the detection of the CPS is possible. In the case of second-order QSPTs, though, like the $E(5)$ CPS, the changes at the critical point are subtle; thus, one may miss the CPS just because the grid provided by increasing the neutron numbers in steps of two is not dense enough in order to achieve coincidence with the critical point somewhere. The introduction of the conformable fractional calculus makes this coincidence more probable since the conformable fractional $E(5)$ CPS contains a free parameter, the order of the conformable fractional derivative $\alpha$. Thus, one has many models similar to $E(5)$, which increases the probability that one of them will fall on the critical point. This result is correct not only at the CPSs but also for all transition legs and dynamical symmetry vertices of the Casten triangle. For each model in the Casten triangle, the conformable fractional BH provides us with many models similar to the original one, increasing the probability that one of them will fall on the experimental results with reasonable accuracy, mainly when the value of $\alpha$ is very close to 1.

In order to prove this result, a comparison between the classical and conformable fractional predictions of two classes of potentials is considered in the present work. We used the finite difference technique for diagonalization of the BH with the Bonatsos potentials, $\beta^{2n}/2$. Conformable fractional predictions for spectra have been given for the potentials $\beta^4/2$, $\beta^6/2$, $\beta^8/2$ and denoted by the $E^\alpha(5)$-$\beta^4$, $E^\alpha(5)$-$\beta^6$, and $E^\alpha(5)$-$\beta^8$ models, respectively. These models are able to close the gaps between the classical spectra of the $U(5)$, $E(5)$-$\beta^4$, $E(5)$-$\beta^6$, $E(5)$-$\beta^8$, and $E(5)$ models entirely. The second class of potentials is sextic potentials. Such potentials are perfect for describing QSPTs and CPSs. This because they may have a single minimum, a flat shape, or two simultaneous minima separated by a barrier, based on their parameters. Many coexisting phenomena in the ground band states can be distinguished depending on the height of the barrier compared to the low-lying energy levels. The experimental evidence that supports these predictions was investigated in the $^{96}$Mo nucleus. The spectrum of the $^{96}$Mo nucleus shows a clear shape coexistence with mixing, as shown by a double-peak structure in the probability distribution. Indeed, the conformable fractional BH provides a framework for a more comprehensive interpretation of the coexisting phenomena.